# Anisotropic Ripple Deformation in Phosphorene

Liangzhi Kou,[*,†] Yandong Ma[‡], Sean C. Smith,[†] and Changfeng Chen[§]

[†]*Integrated Materials Design Centre (IMDC), School of Chemical Engineering, University of New South Wales, Sydney, NSW 2052, Australia*

[‡] *Engineering and Science, Jacobs University Bremen, Campus Ring 1, 28759 Bremen, Germany*

[§]*Department of Physics and Astronomy and High Pressure Science and Engineering Center, University of Nevada, Las Vegas, Nevada 89154, United States*

E-mail: kouliangzhi@gmail.com

## Abstract

Two-dimensional materials tend to become crumpled according to the Mermin-Wagner theorem, and the resulting ripple deformation may significantly influence electronic properties as observed in graphene and $MoS_2$. Here we unveil by first-principles calculations a new, highly anisotropic ripple pattern in phosphorene, a monolayer black phosphorus, where compression induced ripple deformation occurs only along the zigzag direction in the strain range up to 10%, but not the armchair direction. This direction-selective ripple deformation mode in phosphorene stems from its puckered structure with coupled hinge-like bonding configurations and the resulting anisotropic Poisson ratio. We also construct an analytical model using classical elasticity theory for ripple deformation in phosphorene under arbitrary strain. The present results offer new insights into the mechanisms governing the structural and electronic properties of phosphorene crucial to its device applications.

TOC Figure

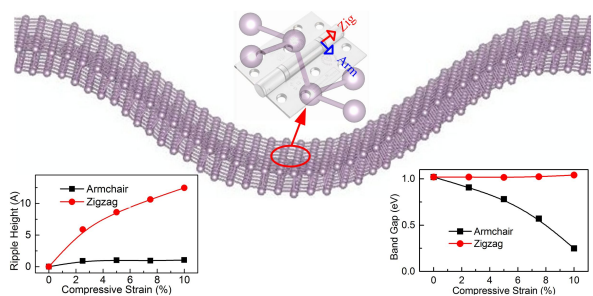

The stability of two-dimensional (2D) layers and membranes is a subject of long-standing debate in materials physics. According to the Mermin-Wagner theorem, [1] long-wavelength fluctuations destroy any long-range order in 2D crystals. Similarly, 2D free-standing membranes have a tendency to crumple.[2] These fluctuations can, however, be suppressed by anharmonic coupling between bending and stretching modes, meaning that 2D membranes can exist but will exhibit strong height fluctuations.[3] The discovery of graphene, a truly 2D crystal, and the recent experimental observation of ripples in suspended graphene,[4] have reignited great interest in these issues. Fundamental insights into the mechanisms of the stability of 2D materials are essential to understanding their electronic properties crucial for device applications.[5-6] Ripple deformation is also present in other prominent 2D layered materials, such as boron nitride[7] and $MoS_2$ monolayers[8-9] where ripples are caused by thermal fluctuations or mismatch strains with supporting substrates.

An important new member of the 2D material family, single- or multi-layered black phosphorus, also known as phosphorene, has been recently fabricated with micromechanical cleavage and exfoliation methods.[10-12] Phosphorene has a significant advantage over semimetallic graphene because it exhibits a finite and direct band gap within an appealing energy range[13-14] and high free-carrier mobility (around 1000 $cm^2$/v.s) [10-11] as well as other unique features, such as anisotropic electric conductance[11,14,15], optical responses[16-17] and negative Poisson's ratio,[18] which distinguish it from other isotropic 2D materials such as graphene and transition-metal dichalcogenides. Given its 2D nature, ripple deformation is expected to spontaneously appear in phosphorene with a significant impact on its electronic properties important to the performance of phosphorene-based nanodevices[10, 19, 20].

In this Letter, we report on a first-principles study on the ripple deformation in monolayer phosphorene under compressive strain. Our calculations unveil a highly anisotropic ripple deformation pattern, where ripple only occurs along the zigzag direction at strains up to 10%, but not the armchair direction. This unusual anisotropic deformation mode stems from a unique puckered bonding configuration of phosphorene. Under compression along the armchair direction, strain energy release is dominated by an angle distortion mode without much ripple deformation; on the other hand, compressive strain along the zigzag direction leads to a predominant bond-compression mode with a steep rise in strain energy, which is released by an out-of-plane ripple deformation. This direction-selective ripple deformation produces an anisotropic response in electronic properties, which is in stark contrast to previous theoretical predictions.[14, 21] We also constructed an analytic model for the ripple amplitude along an arbitrary direction from the classical elasticity theory. The present findings highlight significant effects of anisotropic structural deformation on the mechanical and electronic properties that are crucial for design and operation of phosphorene-based nanodevices.

For the first-principles calculations reported in this work, we used the SIESTA package.[22] The generalized gradient approximation (GGA) was applied to account for the exchange-correlation function with Perdew-Burke-Ernzerhof (PBE) parametrization,[23] and the double-$\xi$ basis set orbital was adopted. It has been shown that the GGA is more accurate than the local density approximation for describing the structural and mechanical properties of phosphorene.[24] During the conjugate-gradient optimization, the maximum force on each atom is smaller than 0.01 eV/Å. A mesh cutoff of 300 Ry was used. Periodic boundary conditions were applied in the two in-plane transverse directions, whereas free boundary conditions were applied to the out-of-plane direction by introducing a vacuum space of 10 Å. To account for ripple deformation in phosphorene, the unit-cell model is insufficient; we constructed a supercell model for this task. Fifteen unit cells (69.27 Å in length) are used for the armchair supercell model, while 20 unit cells (65.99 Å in length) are used for the zigzag supercell model (see Figure 1). The Brillouin zone integration is sampled by a 4×1×10 Monkhorst mesh for the supercell models. Compressive strain is simulated by the reduction ratio of the lattice constant, and an initial out-of-plane ripple deformation is introduced and then the structure is fully relaxed with fixed lattice constants.

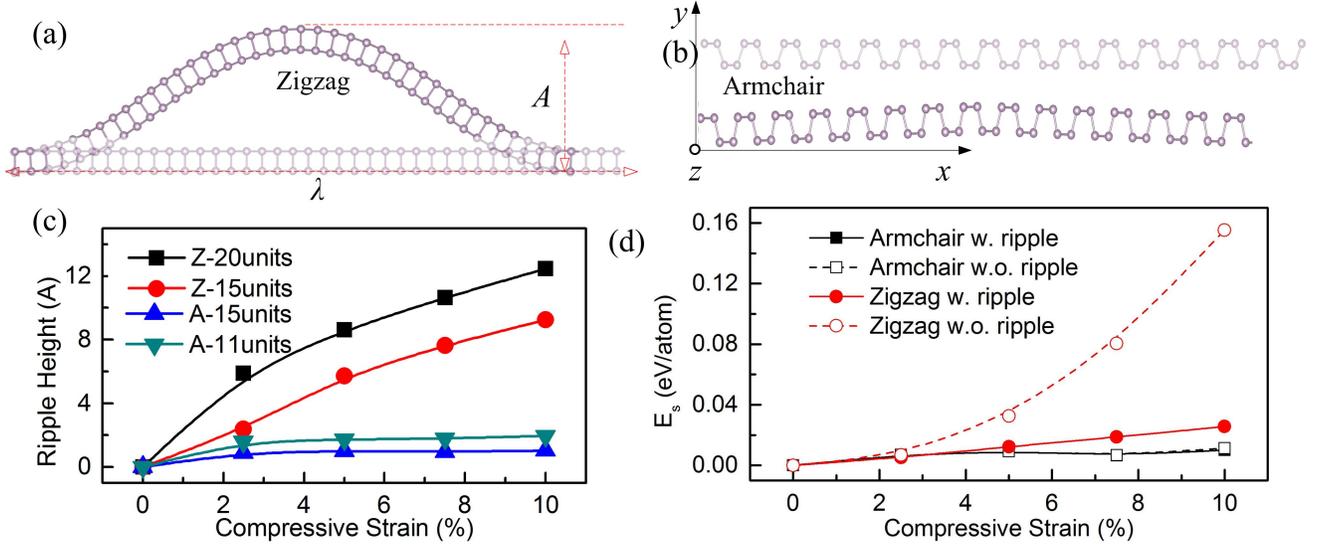

Figure 1: Side view of (a) the 20-unit-cell zigzag and (b) the 15-unit-cell armchair supercell models. The panels correspond to the phosphorene structures under compressive strain of 0% (shade) and 10% (black), respectively. The ripple deformation is characterized by the out-of-plane height $A$ and wavelength $\lambda$. (c) The ripple height versus compressive strain in the armchair and zigzag supercell models. (d) Calculated strain energy for the supercell (with ripple) and unit-cell (without ripple) models of phosphorene under compression along the armchair or zigzag direction.

The optimized lattice constant of the unstrained unit cell is 4.618 Å along the armchair direction and 3.3 Å along the zigzag direction, and the calculated band gap is direct and 1.02 eV at the $\Gamma$ point. These results agree well with those reported in the literature.[11, 14] We show in Fig. 1a and 1b relaxed structures of the zigzag and armchair supercell models of phosphorene under compressive strain of 0% and 10%. The zigzag supercell model develops an out-of-plane ripple deformation under compressive strain, with a shape of sine curve, and the ripple height $A$ increases with strain. In sharp contrast, the armchair supercell model remains almost flat with only a small out-of-plane deformation less than 1 Å up to 10% strain. We plot the calculated ripple heights in the two models as a function of strain in Fig. 1c. It is seen that the ripple height of the zigzag model increases quickly, which can be fit as $A = 220\epsilon - 10^3\epsilon^2$. The ripple shape of the zigzag model under strain can thus be expressed as $y = (220\epsilon - 10^3\epsilon^2)\sin(\frac{x\pi}{\lambda})$. Meanwhile, the out-of-plane deformation of the armchair model remains a small constant after a quick initial rise. These findings suggest highly anisotropic structural response to compressive strain, which is expected to have a major effect on electronic properties. This direction selective development of ripple deformation along the zigzag direction in phosphorene is different from the isotropic local morphology fluctuations observed in graphene.[5] This unusual behaviour stems from the unique bonding structure of phosphorene, as will be further discussed below. To check for possible size effect, we constructed two new supercell models, one with 15 unit cells along the zigzag direction (supercell length $l_0$ = 49.5 Å) and another with 11 unit cells along the armchair direction ($l_0$ = 50.8 Å). Our calculated results (Fig. 1c) reveal the same highly anisotropic structural patters under compression as shown above, namely a pronounced ripple formation (with slightly smaller magnitudes) in the zigzag supercell model and essentially no ripple pattern in the armchair supercell model.

To probe the underlying mechanism responsible for the anisotropic deformation modes in phosphorene under compressive strain, we carried out a comparative study of two sets of structural models for phosphorene in the zigzag and armchair directions, namely the supercell models described above and the unit-cell model used in previous studies. The unit-cell model cannot accommodate ripple deformation, thus can provide information on the bonding changes without ripple deformation and the corresponding buildup of strain energy. We have calculated the strain energy associated with the compression induced deformation, which is defined as $E_s=(E_{strained}-E_{free})/n$, where $E_{strained}$ is the total energy of the strained system, $E_{free}$ is the total energy of the system without strain, and $n$ is the number of atoms in the structural model. The calculated results (Fig. 1d) show that the strain energy of the armchair models increases only moderately with rising strain, and the results for the supercell and unit-cell models are practically indistinguishable. In sharp contrast, the strain energy of the zigzag models increases at a higher rate, especially for the unit-cell model, indicating a strongly strained structural deformation with a high energy cost. This difference in strain energy behaviour is consistent with the previously reported direction dependence of Young's modulus of monolayer phosphorene[26], where the Young's modulus reaches the maximum of 166 GPa along the zigzag direction and the minimum of 44 GPa along the armchair direction.

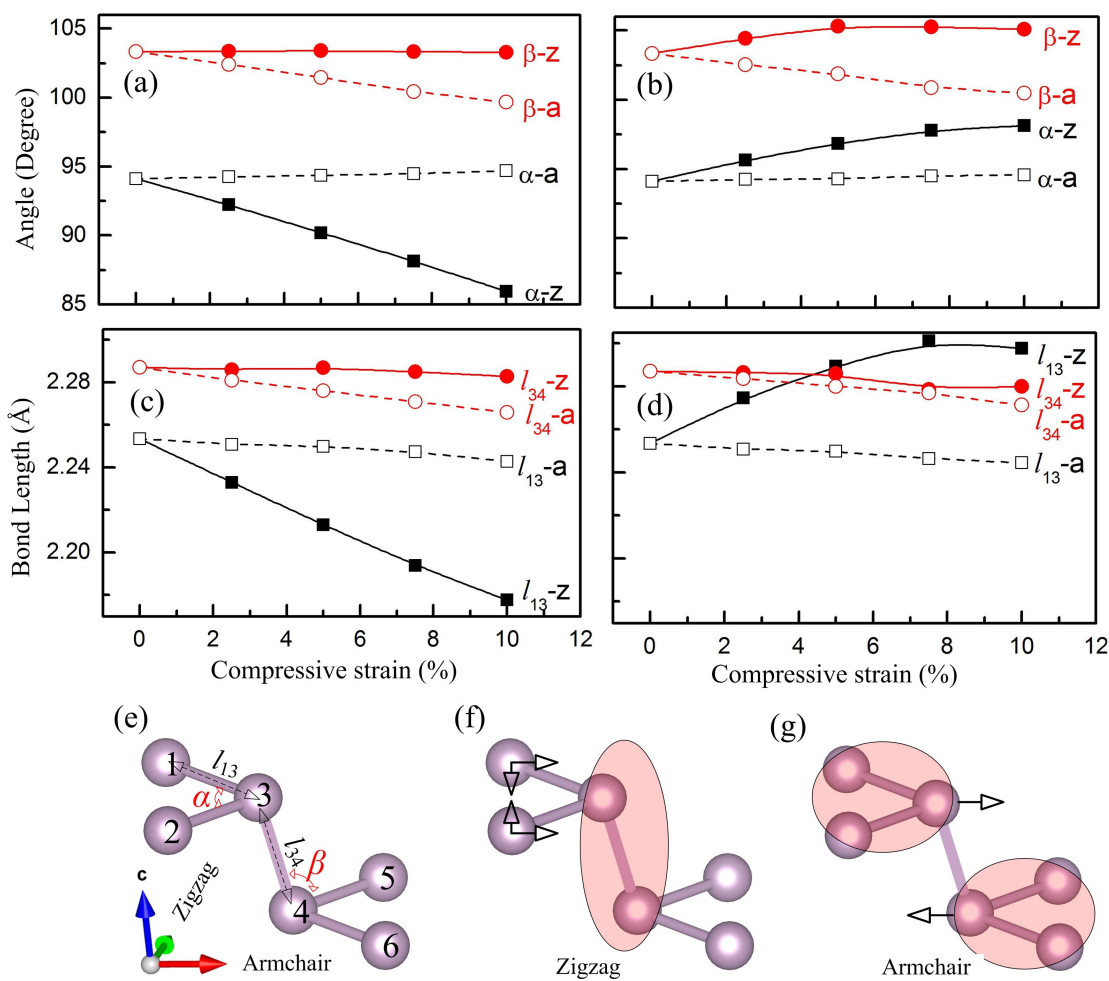

Figure 2: The variation of the hinge angle α and dihedral angle β, bond lengths $l_{13}$ and $l_{34}$ in (a,c) the unit-cell model and (b,d) the supercell model under compressive strains along the armchair or zigzag direction. The structural parameters of the puckered hinge model for phosphorene are defined in (e). The structural responses under compression along the zigzag (f) and the armchair (g) direction are illustrated,

where the shaded areas indicate nearly rigid components during the specified deformation mode, while the arrows indicate the movement of the atoms.

We now examine and compare structural deformation modes in the unit-cell and supercell models to assess the compression induced bonding changes that lead to the highly anisotropic strain energy (Fig. 1d). We track the variation of bonding configurations defined by the following key parameters [see Fig. 2(e)]: (1) the dihedral angle β ($\angle_{345}$), (2) the hinge angle α ($\angle_{132}$), (3) the bond length of the hinges ($l_{13}$), and (4) that connecting the hinges ($l_{34}$). The calculated angle [Fig. 2(a)] and bond length [Fig. 2(c)] variations in the unit-cell model show that the dihedral angle β and bond length $l_{34}$ remain nearly unchanged under strain along the zigzag direction, but the hinge angle α and bond length $l_{13}$ are significantly reduced. Meanwhile, under strain along the armchair direction, dihedral angle β and bond length $l_{34}$ are reduced, but the hinge angle α and bond length $l_{13}$ remain nearly unchanged. These results suggest a puckered hinges model for phosphorene as illustrated in Fig. 2(f) and 2(g). In the zigzag case, the phosphorus atoms in the hinge component move inwards under compression, which reduces the bond length and angles of the hinges, resulting in the large strain energy without ripple as shown in Fig. 1d. This necessitates an effective release of strain energy through the ripple deformation, which is allowed in the supercell model, and this allows the relaxation of both the bond angle and length [see Fig. 2(b,d)], producing a six-fold reduction in strain energy (Fig. 1d). On the other hand, under compression along the armchair direction, the two hinge components (formed by atoms 1-2-3 and 4-5-6, respectively) connected by a nearly rigid rod (the bond between atoms 3-4) are almost free to rotate, and this angular distortion can efficiently accommodate the compressive strain with only a very small strain energy buildup. This is reflected in the nearly identical variations of the structural parameters in the unit-cell [Fig. 2(a,c)] and supercell [Fig. 2(b,d)] model under strain along the armchair direction. Consequently, as shown in Fig. 1d, the strain energies in the unit-cell and supercell models under compression along the armchair direction coincide with each other, thus explaining the absence of ripple deformation. The anisotropic strain energy response unveiled here is consistent and may share the same origin with the recent finding of the interesting mechanical properties and negative Poisson's ratio in single layer black phosphorus. [18, 27]

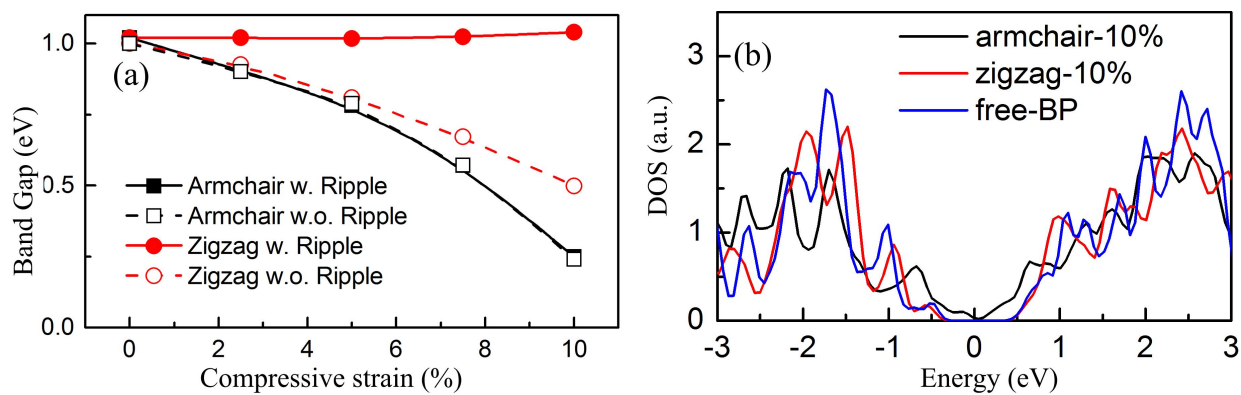

Figure 3: (a) Variation of the electronic band gap of phosphorene under various deformation modes. (b) Density of states under three representative strained conditions.

The highly anisotropic structural deformation in phosphorene, especially the large ripple deformation along the zigzag direction, requires a re-examination of its electronic properties under compressive strain. Here we focus on new trends and behaviour compared to the previously reported results that were obtained using the unit-cell model that does not allow ripple deformation. To this end, we have calculated the electronic band gap of the rippled phosphorene in the supercell model and compared the results with the corresponding data obtained from the strained unit-cell model. Without the ripple deformation, the calculated electronic band gaps of phosphorene in the unit-cell models under compression along both the zigzag and armchair directions decrease almost linearly with increasing strain [see Fig. 3(a)], which is consistent with previous reports.[14, 21] When ripple deformation is considered in the supercell models, the band gap variation shows a strong direction dependence. The results for compression along the armchair direction are nearly identical to those obtained in the unit-cell model, but the phosphorene strained along the zigzag direction exhibits a strain-independent band gap of about 1.02 eV, which is the same value as that of the free-standing phosphorene, in a large range of compressive strain up to 10%. This strain insensitivity of the band gap can be attributed to the ripple deformation which allows an almost full relaxation of the bond angle and length; the compression is absorbed by the out-of-plane ripple, which does not affect the local bonding configuration, thus the nearly unchanged band gap. When strain is applied in the armchair direction, the strain energy is absorbed via angular distortion of the bonds, which leads to substantial changes in the overlap of the wavefunctions of neighbouring atoms, and it produces a significant reduction of the band gap. Our present results correct the previously reported strain modulation of the band gap in phosphorene along the zigzag direction, and these results have important implications for the applications in phosphorene-based flexible nanodevices. The calculated density of states (DOS) in Fig. 3(b) show that the DOS of 10% strained zigzag supercell model closely resemble that of free standing phosphorene, especially near the Fermi level; but the DOS of the armchair supercell model is significantly reduced.

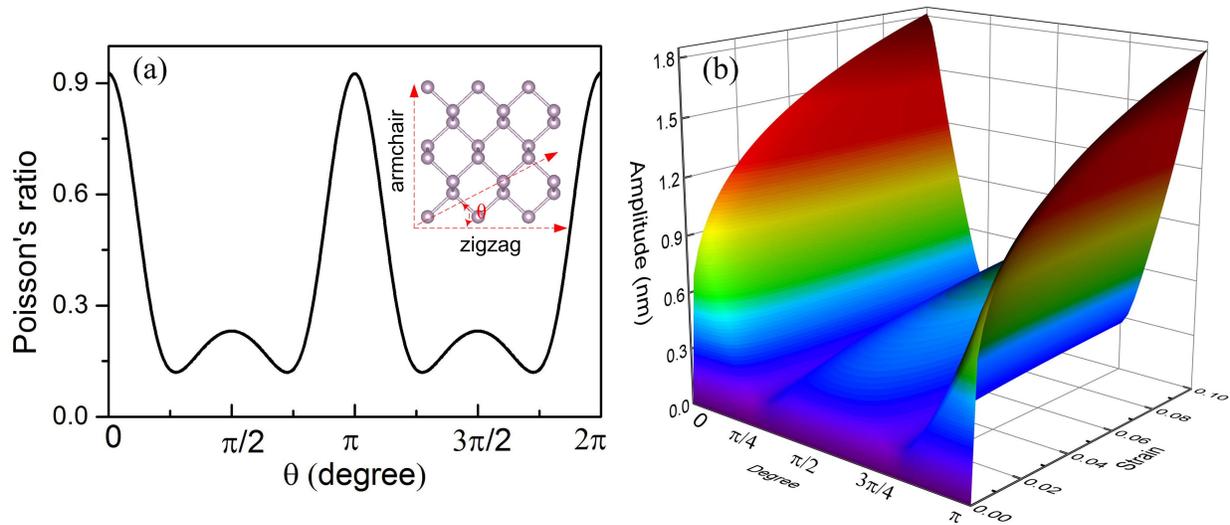

Figure 4: (a) The Poisson's ratio and (b) the ripple amplitude fitted using the classical elasticity theory. The strain direction θ is defined in the inset.

For a more general description of the ripple deformation in phosphorene, we construct an analytical model using the classical thin-film elasticity theory,[6, 28] which establishes the following relation between the longitudinal strain $\epsilon$ in a thin film of thickness $t$ with clamped boundaries at $x=0$ and $x=L$,

$$\frac{A^2}{vtL} = \sqrt{\frac{16\epsilon}{3\pi^2(1-v^2)}}$$

where $A$ is the out-plane amplitude and $v$ is the Poisson's ratio, which is known to be anisotropic and direction dependent for phosphorene. We derived the expression for the Poisson's ratio of phosphorene $v(\theta)$ along an arbitrary direction θ by its stress-strain relationship $\sigma = C_{ij}\epsilon$ ($C_{ij}$ is stiffness tensor) as (see Supplemental Information for more details)

$$v(\theta) = \frac{v_z \cos^4\theta - k_1 \cos^2\theta \sin^2\theta + v_z \sin^4\theta}{\cos^4\theta - k_2 \cos^2\theta \sin^2\theta + k_3 \sin^4\theta}$$

Here $v_z$ is the Poisson's ratio along the zigzag direction, which is 0.93,[18] $k_1 = \frac{C_{11}+C_{22}-\frac{C_{11}C_{22}-C_{12}^2}{C_{33}}}{C_{22}}=1.591$, $k_2 = \frac{2C_{12}-\frac{C_{11}C_{22}-C_{12}^2}{C_{33}}}{C_{22}}=-0.806$, $k_3 = \frac{C_{11}}{C_{22}}=3.259$ (the values of $C_{ij}$ are from experimental measurements[29] and previous theoretical predictions[26]). Our calculated results in Fig. 4(a) show that the Poisson's ratio along the armchair direction (θ =π/2) is only 0.3, which is significantly smaller than the value along the zigzag direction (0.93 at θ =0, π). This conclusion is consistent with previous reports [18]. Combining above two equations, and taking the thickness t=0.56 nm and L=10 nm, we have obtained the out-plane ripple amplitude under strain $\epsilon$ along an arbitrary θ as shown in Fig. 4(b). It is clear that the ripple deformation is most pronounced when θ falls in a narrow range near 0 or π, namely when the applied strain is along the zigzag direction. The out-plane amplitude increases quickly as a quadratic function of strain $\epsilon$. At 10% strain, the ripple height is 18 Å; in contrast, the ripple deformation along any other direction (especially the armchair direction with θ= π/2) regardless of the applied strain magnitude is minimal (below 3 Å). These results of the analytical model in classical elasticity theory are consistent with those of the first-principles calculations shown above. These findings show that in addition to its successful application to graphene layers[6], the classical thin-film elasticity theory also describes well the highly anisotropic ripple deformation in phosphorene, indicating its applicability to atomically-thin 2D membranes in general.

In summary, we show by first-principles calculations that phosphorene exhibits interesting and unique anisotropic ripple deformation behavior under compressive strain. Significant ripple patterns only form along the zigzag direction, where the out-of-plane deformation effectively releases the strain energy. In contrast, compression induced deformation along the armchair direction is dominated by bond-angle distortion without any appreciable ripple formation. The direction selective ripple behavior in phosphorene stems from its puckered structure, which can be regarded as a coupled two-component hinge system. The special alignment of the local bonding configuration leads to the distinct deformation modes. This highly anisotropic structural change produces equally anisotropic response of electronic properties under compression. The striking anisotropic ripple deformation behavior in phosphorene is well captured by an analytical model from the classical elasticity theory. The present findings provide crucial insights into the anisotropic band gap modulation in phosphorene essential to its applications in nanodevices, such as microelectromechanical systems and flexible electronic devices.

## Acknowledgement

We thank Chun Tang for useful discussions. L.K. acknowledges financial support of UNSW Australia's SPF01 project 6301 (CI Smith). C.F.C. was partially supported by the DOE through the Cooperative Agreement DE-NA0001982.

## Supporting Information Available

The detailed derivation of Poisson's ratio is presented. This material is available free of charge via the Internet at http://pubs.acs.org/.